\documentstyle[12pt,aaspp4]{article}

\begin{document}
\include{psfig}

\def \cmm  {cm$^{-2}$}
\def \kms {km~s$^{-1}$}
\def \Lya {Ly$\alpha$}
\def \Lyb {Ly$\beta$}
\def \Lyg {Ly$\gamma$}
\def \Lyd {Ly$\delta$}
\def \th {$\tau_{\rm{HeII}}$ }
\def \zabs {$z_{abs}$}

\def \HI   {\ion{H}{1} }
\def \HeII {\ion{He}{2} }
\def \CIV  {\ion{C}{4} }

\def \object {Q0014+813}

\title{New Keck Spectra of Q0014+813:
annulling the case for high deuterium abundances}

\author{DAVID TYTLER \altaffilmark{1}, SCOTT BURLES\altaffilmark{2}  and 
DAVID KIRKMAN\altaffilmark{3}}

\affil{Department of Physics, and Center for Astrophysics and Space Sciences \\
	University of California, San Diego \\
	MS 0424, La Jolla, CA 92093-0424}

\altaffiltext{1}{tytler@ucsd.edu}
\altaffiltext{2}{scott@cass154.ucsd.edu}
\altaffiltext{3}{dkirkman@ucsd.edu}

\begin{abstract}

The spectrum of 
quasar 0014+813 was the first in which deuterium was claimed to be
observed, at first as a probable detection, and later ``confirmed''.
It is important because it is the only case in which there is a serious 
likelihood of a high D/H ratio, since the others are more ambiguous and
admit a wide range of D/H including zero.
We present new high-resolution spectra with higher
quality than those published.
Deuterium has been reported in two Lyman limit systems (\zabs = 3.32, 2.80)
along the line of sight. In both 
we find that the absorption is readily explained by H alone and that
deuterium absorption is neither required nor suggested by the spectra.
The three previous models with high D/H values
(D/H $\approx 20 \times 10^{-5}$) are not compatible with
the new data, and in some cases the differences are surprisingly large.
Since the spectrum is compatible with all values of D/H from zero up to
large upper limits, neither absorption system can be used to measure D/H, and 
previously cited values should be disregarded.
There is now no case in which a high value of D/H has been established, and all
data are compatible with low D/H. Deuterium has been seen and
measured in only two QSOs (Q1937--1009 and Q1009+2956), both of which give low 
D/H. Several other QSOs give upper limits which are about ten times less 
sensitive.

\end{abstract}

\section{HISTORY OF Q0014+813}

Q0014+813 is a bright (V = 16.9), high redshift ($z_{em} = 3.37$) quasar,
which has been well studied since its discovery in 1983 (Kuhr et al. 1983).
Measurements of D/H have been claimed in two absorption systems towards
this QSO. The first, at \zabs = 3.32 has been discussed in three observational
papers, and is by far the best case for high D/H.  This Lyman limit system is 
also known to be very metal-poor (\cite{cha85}, \cite{cha86}).  

Songaila et al. (1994, hereafter SCHR)
observed this QSO with the Keck-1 telescope and the HIRES spectrograph.
They found an absorption line near the expected position of deuterium,
but could not constrain the model fit, and only
reported an upper limit, D/H $< 25 \times 10^{-5}$.
The portion of their spectrum around D (their Figure 3) appears to be 
smoothed, as can be seen from the strong correlations in the counts of
adjacent pixels. They do not mention smoothing and quote the nominal HIRES
resolution of R=36,000. Rugers \& Hogan (1996a, hereafter RH) state
that SCHR did smooth the data.

Carswell et al. (1994, hereafter CRWCW) analyzed the same absorption
system with lower SNR spectra from the echelle spectrograph on the KPNO
4m telescope.  They showed that many different models could be
applied to their data, and found D/H $ < 60 \times 10^{-5}$.
The models of SCHR \& CRWCW
are consistent with earlier studies by Chaffee et al. (1985, 1986).
In particular, there was excellent agreement on the total
hydrogen column density, both SCHR and CRWCW gave Log N(H~I) = 16.9 \cmm, and
Chaffee et al. (1985) gave Log N(H~I) = 16.8 \cmm.

RH re-reduced the SCHR spectrum using the standard IRAF package.
They did not smooth the data, and saw a 
strong ``spike'' (3 pixel region of increased counts) in the deuterium 
absorption feature.
They considered this real, since the top of the spike was 6 -- 7 $\sigma $ 
above the bottom, and it was 
also seen when the data were reduced using the SCHR software.
They devised a new model with two D components, one on either side of the
spike. The H must then also be split into two. They increased
the total N(H~I) column by 70\% to Log N(H~I) = 17.2 \cmm, outside the ranges
of both previous studies.
RH reported that the new model gave a direct measurement
of D/H in both components, which is surprising, because the complexity and 
number of free parameters was higher than in all previous models.  They found
D/H $= 19 \pm 5 \times 10^{-5}$ in each component.  

In this paper, we present new HIRES spectra of Q0014+813, which has
the highest resolution and SNR to date.  We compare our data and model fits
with those of SCHR, CRWCW \& RH and show that D/H cannot be measured
in this system.  

Rugers \& Hogan (1996b) noticed that a second higher column density system 
with Log N(H~I) $> 18$ \cmm, at
lower redshift, $z_{abs} = 2.80$, contained unusually narrow and unresolved
metal lines. They obtained a measurement of D/H with only the blue
wing of the \Lya~ line where D was not even resolved.  They argued that the
steepness of the line required a fit with D.  We show that this is
incorrect. The fit with D is not unique and cannot constrain D/H.

\section{OBSERVATIONS \& DATA REDUCTION}

On the nights of Nov 2nd \& 3rd, 1996, we used the HIRES
spectrograph (Vogt 1994) on the Keck-1 10-m telescope to obtain spectra
of Q0014+813 in four separate exposures totaling 4.5 hours. 
Each of the four spectra provided wavelength coverage of 3700 to 6100 \AA,   
although the useful portions of the spectra only extend down to about 3900 \AA.
The first three exposures were taken with a 0.8" x 5.0" slit, 
which gave a resolution of 6 \kms.  With the fourth exposure, we used a wider
slit (1.45" x 5.0") to reduce slit losses, which gave a 
resolution of 10 \kms.  
The newly commissioned HIRES rotator (Tytler et al. 1997)
was used during all observations
to minimize losses due to differential atmospheric diffraction; losses which
can be substantial below about 4000 \AA.  
The spectra were optimally extracted with the automated program described
in Burles \& Tytler (1996). 
For wavelengths above 4100 \AA, we added only the 3 higher resolution spectra 
and not the fourth one, which would decrease the spectral resolution.
Below 4100 \AA, the lower resolution spectrum was coadded to increase
the SNR.  First we convolved the higher resolution spectra with a gaussian to
match the low resolution spectrum, and then we coadded all 4 spectra.
The final spectral resolution, which we need to find intrinsic line 
parameters, was measured from the arclamp line widths.
All wavelengths are vacuum  
values in the heliocentric frame.

\section{MODEL FITS}

There are several ``rules'' which we use to assess fits.
We discuss model fits of Lyman limit systems and other D/H techniques in more 
detail in Tytler \& Burles (1996).

(1) Model fits do not need to explain all of the absorption in a region of the
spectrum, since there is often additional unrelated H absorption, which may
have no effect on the parameters of interest.
(2) Fits must never absorb flux which is seen.
This would be unphysical, and would immediately rule out a fit, provided the
``flux'' was real and not noise.
(3) The best model is the one which uses the least number of parameters to 
fit the data. This usually means the least number of velocity components.
(4) A model fit should give a reasonable chi-squared statistic
($\chi^2 \approx 1$) if all absorption in the spectrum is 
included in the model.
(5) Noise can be judged from the fluctuations in the bottom of the
saturated H lines, which should have exactly zero flux when there is no
noise, no sky subtraction errors and no stray light.
(6) When the signal to noise is high, we need full page plots which show the 
noise level and each pixel, to see if fits are acceptable.  
Fits which differ from the data by 5\% are rejected with SNR  =100.
(7) A complete model would account for all data on an absorption system. 
(8) All lines and continuum absorption from a given ion should be fit with the
same parameters ($b$, N and $z$).
(9) Different ions may require different parameters ($b$, N and $z$).
(10) Wherever we see metal lines we expect to see H lines, but the parameters 
($z$, $b$) may appear to differ if there are hidden unresolved components eg.
with different metalicity and ionization.
(11) The absolute values of line widths should be physically reasonable,
e.g. $b > 15$ \kms~for Log N(H~I) $>16$ \cmm.
(12) The ratio of $b$ values should be physically reasonable. The line
widths arising in a single absorber should 
be fully described by a temperature and a turbulent velocity.
(13) A line is unlikely to be D if we find that we can not fit it using only
D plus the blend with its own H. 
When additional absorption (e.g. H at another velocity) must 
be introduced, it is generally possible to fit the whole line with this new H.
We can then have any D/H from zero up to some upper limit.

\section{ABSORPTION SYSTEM at $z_{abs} = 3.32$}

Figure 1 shows a region of the spectrum containing the saturated
\Lya~ absorption 
feature at \zabs = 3.32.  
The absorption line at 5251.45 \AA, identified by 
SCHR, CRWCW \& RH as possible D absorption, is marked by the
left tick mark in Fig. 1.   The position of the associated H
absorption is shown with the right tick mark near 5252.88 \AA . 
To determine if the line at 5251.45 \AA~ is D,
we start with the following:
the line has a column density of Log N = 13.50 $\pm$ 0.03 \cmm, a velocity 
dispersion of
$b=24 \pm 1$ \kms, and a redshift of $z=3.320976$.  For the rest of this
letter we will refer to this redshift as $v=0$ \kms, and refer to all
other line positions as velocity displacements from this redshift.
The position of this line is very well determined and optimally fit with a
single component at
$z= 3.320976 \pm 0.000007$ ($\pm 0.6$ \kms).

We must now determine the properties -- most importantly the
positions -- of the H~I absorbers.  The \Lyb,\Lyg, \& Lyd transitions show
that there are at least two components (see Figure 2), which we refer
to as the blue (A: $v \simeq 10$ \kms) and
red (B: $v \simeq 120$ \kms) components.
These components are well separated, by about 108 \kms. They are not the
close components considered by RH which were both near 0 \kms.
For a D/H measurement the only
important information comes from the absorber or absorbers in the blue
component, because the D from the red component would be in the bottom of the
saturated \Lya~ line where there is no flux.  

The blue side H absorption at $v \simeq 0$ can be fit reasonably with
one H~I absorption component (A in Fig. 2a), 
That absorber has Log N(H~I) = 16.7 \cmm, $b=23.8$\ kms, and $v=10 \pm 0.5$
\kms. The large velocity offset is the main reason why D alone can not
explain the absorption feature at 5251.45 \AA . There must be significant 
additional absorption, e.g. by H. 
For this reason, we conclude that if the
blue side H~I absorption contains only one component, this system is
unsuitable for a D/H measurement. At best we get a high upper limit, which
still allows D/H $=0$.

Table 1 shows goodness-of-fit values for each model presented.
We show reduced $\chi^2$ values over critical regions in the spectrum.
The best fit is found by minimizing the total reduced $\chi^2$
of all the Lyman lines.
Fits are acceptable when the reduced reduced $\chi^2$ is about one.
In Table 1, we show only the regions that are most sensitive to
the position and column density of D.
The 1 component model is particularly bad on the blue side of H~I \Lya~around
--40 \kms~ $<v<$ --60 \kms.

A much better fit is obtained with two components
(B and C in Figure 2b), which is the maximum number justified by
this spectrum.
The stronger H~I component (C) is at
$v=17 \pm 1$ \kms, with Log N = $16.61 \pm 0.05$ \cmm~
and $b = 19 \pm 1$ \kms, and gives a better fit to the high-order
Lyman lines (i.e. Lyman 13-17). 
With component C in this position, the second component B
is needed to fit the low order lines. 
Because the best 
evidence for component B comes only from saturated regions of
absorption (\Lya~ through \Lyd),  we find that it 
is not well determined -- many different
N, $b$, and $v$ values for component E give a good fit to the data.
The best fit occurs for Log N = 16.12 \cmm, $b = 17$ \kms, and $v=-13$ \kms.
Once again, because both of the H lines are far from the $v=0$ required to fit 
D, the feature at 5251.45 \AA~ can not all be D.
In Figure 2b we model
all of the 5251.45 \AA~ feature as a single unassociated H line
(component H) with 
Log N(H~I) = 13.52 \cmm, $b$ = 24 \kms, $z$ = 3.319799, and $v =$ --81 \kms.
The reduced $\chi^2$ on the blue side of \Lya~ is $\approx 4$,
which indicates that the model fit could still be improved between
--40 to --60 \kms.  In Ly-10 \& Ly-13, the reduced $\chi^2 \approx 2$ 
results from extra \Lya~ absorption which we have not included in the model,
This unrelated H~I absorption does not affect the results and would only
complicate the models presented here.

Figure 2c shows the best fit which we can obtain if the 
core of the 5251.45 \AA~ feature is the D of component E. 
Here we have moved component B (Fig. 2b) to a higher velocity to better match
5251.45 \AA~ .  This fit is unacceptable 
because the blue side of \Lya~ gives $\chi^2 \, = 11.9$. 
It is also an unusual fit because the D absorption 
is strongest in the --4 \kms~ component (E),
while H is strongest in the +16 \kms component (F), so
that the D/H values in the two components to differ by a factor of 30.
The feature is fit by a blend of D absorption
from E and F, and 
because there are many possible values of $v(E)$, we can not measure D/H
in either component.
Even if move component E around to best fit 5251.45 \AA, we do not get 
acceptable fits because the same gas must explain the blue side of Lya~
(at --40 to --60 \kms).

Regardless of how the blue side H~I absorption is fit, as one
component or as two, the 5251.45 \AA~ absorption feature requires the
presence of interloper absorption.  Because an interloper {\it can} explain
all of the absorption, and we know that one {\it must} be present, no
measurement of D can be made in this system. 
Because we have the highest SNR observations to date of this object,
we conclude that all previously published D/H measurements in this
object are incorrect.  We also conclude that no measurement of D/H in
this system can be made in this system.

\subsection{Other published models of \zabs = 3.32}

Three previous models have been published for this absorption system,
and we have discussed them briefly in I.  The first 
two studies concluded with best fit
values of D/H with no formal limits (SCHR \& CRWCW), and 
the third (RH) claimed to measure the same D/H in each of two components.
We now compare the best models form each study with the new spectrum.

There are considerable differences in the continua levels. In Figure 4
we show our continuum level fit to one echelle order.
We see that a slowly changing smooth (third order) function gives an 
excellent fit to the high points, which are known from many other studies to 
be the continuum.  There is no need for  small scale irregularities.
The continua in two other studies can be deduced by tracing the normalized
flux published in each study and applying it to our spectrum.
For example, at every pixel where the normalized flux is unity in a 
published spectrum, the corresponding continuum level in our spectrum
must equal the flux in that pixel. 
The SCHR continuum is concave and drops below ours by 14\% , 
they assume little or no absorption in the peaks 
between the lines near 5246, 5257 and 5259 \AA.
CRWCW continuum is approximately constant and low by about 9\%. This looks 
like a reasonable continuum based on the restricted wavelength
coverage they present in their Figures 1 \& 2. 
Although both continua have surprisingly large errors, the continuum
level is not as important as the velocities.

There are also large differences in velocities.
CRWCW find D at $v = +8$ \kms, a good agreement
with our wavelength scale and spectrum because this is an ``external'' absolute
error.  SCHR quote D at $v = -75$ \kms.
Most of this error is removed if we assume that they forgot to correct
their data to vacuum wavelengths. With this correction their D is at +8 \kms.
If they did not apply a heliocentric correction then their
D would be at +4 \kms.
We do not understand the wavelength calibration of RH, but we find
a mean velocity difference of $v = -23$ \kms.
Since all three papers have zero-point offsets in the wavelength scale, 
we now apply small velocity shifts to give their models the best fit
to our data.

SCHR use a five component model, with only one strong blue H~I component
at $v = +4$ \kms~ (Fig. 3a).  The H~I component is
too red in \Lya, \Lyb, but too blue in Ly-13.  This model uses 
D/H = 25 $\times 10^{-5}$, and assumes b(D~I) = b(H~I) / $\sqrt{2}$.
The amount of absorption is no where near sufficient to account for the
whole feature at 5251.45 \AA, so we must conclude that interloping
absorption is necessary for the SCHR model to give an adequate fit.

CRWCW also model the H~I with one strong blue component, but they add two
weak H~I absorbers near the D position (this absorption is blended
with the D line).  The fit by CRWCW is by far the best fit of the 3 different
prior studies, even though they had much lower quality
data than SCHR \& RH.  CRWCW made very conservative conclusions, which we
confirm: {\it We reiterate that no single measurement ... should be regarded as
providing anything other than an upper limit... The results here should
be confirmed (or not) and improved by obtaining and analyzing high S/N,
high resolution spectra}.  We show only one of the models proposed by CRWCW,
with D/H =$24 \times 10^{-5}$ (see Fig. 3b). Their
full range of models allowed D/H as high as 60 $\times 10^{-5}$.

RH model the H~I with two strong blue components with positions 
separated by 21 \kms.  The mean velocity position of the two components
is --1 \kms.  The two components were demanded not by the
structure of H~I in the high-order Lyman lines, but by the spike in the
spectrum which split the deuterium feature in two.  This spike
left two very narrow $b < 9$ \kms~ lines, which were they assumed
{\it a priori} to be deuterium.  We have four separate spectra of Q0014+813,
and this spike appears in none of them.  RH stated that the spike must
be real because it also appears in the data of SCHR.  
Our higher quality spectra show that the spike is not real. 

Amazingly RH were able to make two {\it independent} measurements
(each accurate to 30\%) of D/H which
agree exactly.  RH analyzed the same data 
that allowed SCHR to quote only a 
best guess.  Figure 3c shows the dual component model proposed by RH, which
is nothing like the data (see Table 1).
In addition, the H~I column
density was overestimated by more than 50\%. The fit to
Ly-13 shows that this is a very major error, which is easy to see.  
Tytler \& Burles (1996b) have previously noted this problem:
the RH fit could be seen to unphysical 
because it absorbed flux where light was seen in their Figure 2.
To compensate for their overestimation of N(H~I), and to allow a fit to \Lya,
RH required extremely low velocity dispersions for H~I. 
One blue component is modeled with $b=10.1$ \kms, which corresponds
to a cloud temperature of $T < 6500$ K.  No \Lya~ absorber with 
Log N(H~I) $\approx$ 17 \cmm~
has ever been measured with such a low velocity dispersion.
The RH fit to this system is unphysical and seems to have been designed to
give a high D/H fit.

There are at least two possible origins for the ``spike''.
SCHR applied a  filter to remove cosmic rays (CRs), but we see four large
spikes in the top panel of their Figure 1 (5224, 5232, 5239, 5256), which 
look like CRs, and might account for the spike at the D position.
RH used the IRAF IMSUM procedure, which they note left some CRs.
There are also hot pixels on the HIRES CCD, each of which should be 
flagged because they are not removed or revealed by normal data reduction.

\subsection{Conclusions on $z_{abs} = 3.32$}

We conclude that most of the H~I is at 9 \kms~ $< v <$ 18 \kms, and
not at $v=0$ where it should be if the 5251.45 \AA~ feature is D. 
If there is D in 5251.45 \AA ~then at least one additional 
line is needed to give a fit: either an interloper if H is one
component, or D from component F if H is two component).  
There is no way to
determine the position of this additional line, so the
fit is ill-constrained and no
D/H measurement can be made.  Only weak upper limits can be placed 
on D/H in each component. 

If we try to fit all of 5251.45 \AA ~using D at the velocities of the two 
components (E \& C) which best fit the H, then 
we get the result shown in Figure 2d, which is
completely unacceptable, again because the velocities do not fit.
But this models allows us to place 
2$\sigma$ upper limits 
on D/H of $ < 100 \times 10^{-5}$ and
$< 35 \times 10^{-5}$ for components B \& C, respectively. 
The fit with these upper limits is shown in Figure 2d.

\section{ABSORPTION SYSTEM at $z_{abs} = 2.80$}

The spectrum we presented above also covers another Lyman limit
system (at \zabs = 2.80), in which another measurement of D/H has been made
and published.
Rugers \& Hogan (1996b) analyzed this system with the same spectrum presented
in RH.  They concluded that the blue wing of \Lya~ must be D and found
D/H = 19 $^{+16}_{-9} \times 10^{-5}$.  Our new spectrum has the 
same wavelength coverage of the spectrum studied by Rugers \& Hogan, so we
can test the goodness and uniqueness of their 
model fits on our higher quality spectrum.

This system has a high neutral hydrogen column density, 
Log N(H~I) $> 18$ \cmm, determined by the opacity at the Lyman limit
and the damping wings of \Lya.  
The hydrogen column density is divided among many components.
The strongest of these components have accompanying metal lines which
allow accurate determinations of their velocity positions.
The weaker hydrogen components will not usually have significant metal lines
and then their positions can only be determined with the lines in the Lyman
series.  
This presents a serious problem: because of the Lyman limit
system at higher redshift and a steep decline in sensitivity as wavelength
decreases, only the spectral region covering \Lya~ can be used to place
constraints on the model for this system.  Therefore, any information
on the column densities and velocity dispersions of D and H 
must come from the \Lya~ feature alone.

The top panel of 
Figure 5 shows the large saturated \Lya~ feature at $z_{abs} = 2.80$.
The tick marks indicate the velocity positions of the strongest components
determined by the accompanying
metal lines.  Figure 5 also shows 6 of the metal line complexes associated
with this system.  Our data require at least 10 components for an
acceptable fit to the metal lines (shown by the solid curve).  
The high number of metal line components
introduces at least 20 free parameters to the model fit of \Lya.  

Many of the metal lines have narrow intrinsic
velocity dispersions, b $< 5$ \kms, but the components are spread
over a range of velocity, $\Delta V = 300$ \kms.  The large velocity
range is much greater than the velocity splitting of D and H (81.6 \kms),
Only in the lowest velocity component (bluest component) could deuterium
be detected.  The bluest component has low velocity dispersions in the
metal lines, b(Si) = 4.1 $\pm 0.5$ and b(C) = 4.8 $\pm 0.5$.  The low
values imply an upper limit to the temperature in this component,
T $< 22,000 K$, and also upper limits to the velocity dispersions of
H and D: b(H) $< 19$ \kms, b(D) $< 13.4$ \kms.

Rugers \& Hogan argue that the blue wing can only be fit with a very narrow
line, too narrow for hydrogen.  Our data show that this claim is
false.  If the line is placed at the expected position of D
corresponding to the bluest metal component, then it must be
(1) narrow, and (2) not completely unsaturated (see Figure 6a).
The maximum line width is set by the metal lines, and the line can not be
made narrower without reducing its N(D~I) and hence its central depth.

However is this line is H it can be
placed freely, and an excellent fit is found with an
ordinary H line with Log N(H~I) = 14.1 \cmm~ and b = 21 \kms~ (see Figure 6b).
The single H line provides a good fit to the data, and it lies
16 \kms~ redward of position expected for D in the H associated with the blue
most metal component.

It is tempting to place an upper limit on D/H in the bluest component,
but this would be a dangerous exercise.  The D line accompanying the
bluest metal component has a strong upper limit,
Log N(D~I) $< 13.8$ \cmm, but the 
column density in the corresponding H line has no lower
limit.  The model of Rugers \& Hogan constrains the bluest component
to a column density, N(H~I) = 18.04 $\pm 0.12$ \cmm.  The constraint
comes from the damping wings of \Lya, but this fit is not unique and
the data do not require a column density in this range.
The column density in individual components is difficult to determine
in saturated features, even with high SNR data
(Tytler, Fan \& Burles 1996; Wampler 1996).
An upper limit on D/H requires a lower limit on N(H~I), which can only
be measured in the high-order Lyman lines.

We conclude that it is impossible to measure D in this system
with only spectral coverage on \Lya.  The D line must be resolved from
the H feature to rule out contamination from other H lines. 
The model presented by Rugers \& Hogan is ruled out at the 10$\sigma$
level in our data, and cannot be considered a measurement of D/H.  

\section{CONCLUSIONS}

New high resolution, high SNR spectra
of Q0014+813, show the following:

1. Measurements of D/H cannot be made in either of the two Lyman limit
absorption system towards Q0014+813.  

2. The candidate D line at \zabs = 3.320976 cannot be properly fit
without introducing contamination from other lines.  The exact amount and
nature of the contamination cannot be constrained and precludes any measurement
of D/H. 

3. The strong \Lya~ absorption at \zabs = 2.80 does not require 
D absorption, and gives the best fit in models with no deuterium.  
\Lya~ is the only Lyman line which constrains the column densities
of the H components.  Without high SNR spectra of the other Lyman lines,
we cannot even place upper limits on D/H in this system.

4. The only published measurements of high D/H are towards
Q0014+813.  The analysis we present here removes any suggestion of high
D/H in QSO absorption systems.

5. Deuterium has been seen only twice in QSO absorption systems.
In each case, we obtained a measurement of D/H (not an upper limit), because 
the D line
can be fit by the velocities determined from metal lines, the D line widths
agreed with predictions from the metal and H lines, and we applied
statistical corrections for contamination by H.
The first measurement of D/H in a QSO absorption system was made towards
Q1937--1009 (Tytler et al. 1996), and the second towards
Q1009+2956 (Burles \& Tytler 1996). 
Both measurements give low values of D/H, which agree to within 15\%.
All other cases are upper limits, because of either poor data, or
complex absorbtion systems which cannot yield strong constraints on D/H.
All data in QSO absorption systems are fully consistent with low D/H.

\clearpage

Figure 1:  The \Lya~ feature at \zabs = 3.32 in the Keck HIRES spectrum
(FWHM = 6 \kms).  The left tick marks the centroid of the 
line at 5251.45 \AA and the candidate for D absorption.   The right tick
marks the position of the associated H line.  We set these positions to
zero velocity for the remainder of the analysis.

Figure 2: The Lyman lines of the \zabs = 3.32 system.  In all of the plots
the tick marks correspond to the line centers of individual components used
in the model fits.  We present 4 different fits to the data to investigate
the possible models which could give an adequate measure of D/H.  None
of the models succeed in giving a measurement.  The SNR in each panel
is 55, 20, 17, 13, and 10 per 2 \kms~ pixel for \Lya, \Lyb, \Lyg, Lyd, and
Ly-10 and above respectively.  See the text for the descriptions and
interpretations of each model.
For Ly-10 to 17 we see absorption from more than one series line in each
panel: e.g. velocity --40 \kms~ in Ly-16 is also +66 \kms~ in Ly-17. 
Each tick marks the position of a hydrogen or deuterium Lyman line.
In the higher order lines, the components of the Lyman series begin
to overlap, and the model profile becomes more complicated.  

Figure 3:  Four of the Lyman lines shown in Fig. 2.  The models shown
in these 3 plots are from other published studies of this system.
We overlay the fits on our higher quality spectrum to provide
a direct comparison of the different models.  In Table 1, we quantify
the goodness-of-fit of each model to the spectrum.

Figure 4: Comparison of Continuum Levels. The \Lya~ feature at \zabs=3.32
is centered near 5254 \AA.
The histogram shows counts from one of our four new spectra.
These are ``raw'' counts, which have not been flux calibrated.
Our continuum is the solid smooth curve which lies high on the data, because
there is a lot of \Lya~ absorption at $z > 3$.
The dashed line
shows the fit that would be required to give the normalized spectrum
in SCHR.  The dotted line is an approximation to the fit required to
give the normalized spectrum in CRWCW.

Figure 5:  The \Lya~ and metal lines at \zabs=2.80.  Zero velocity is
set at the position of the bluest metal line.  The fit to \Lya~ was produced
from the model of Rugers \& Hogan (1996b).  

Figure 6: (a)  The model fit of \Lya~ at \zabs=2.80 
with an upper limit on N(D~I).  The lack
of a lower limit on the corresponding H~I column density prohibits any
useful constraints on D/H in this system.  
(b) An extra H line gives a very good fit to the data and requires no D,
and it does not 
fall at the position of D predicted by the bluest metal line.

\clearpage

\begin{table*}
\begin{center}
\begin{tabular}{ccrrr}
Model & Figure & $\chi^2$(\Lya)\tablenotemark{a} 
& $\chi^2$(Ly-10)\tablenotemark{b} 
& $\chi^2$(Ly-13)\tablenotemark{b} \cr
\tableline
1 Comp & 2a & 6.54 & 2.30 & 2.59 \cr
2 Comp & 2b & 3.87 & 2.29 & 2.14 \cr
2 Comp (force D) & 2c & 11.9 & 2.13 & 2.43 \cr
2 Comp (with D) & 2d & 75.2 & 2.29 & 2.14 \cr
SCHR & 3a & 171.6 & 4.04 & 5.77 \cr
CRWCW & 3b & 22.4 & 6.31 & 4.98 \cr
RH & 3c & 45.9 & 6.42 & 16.0 \cr

\tablenotetext{a}{measured over --100 \kms~ $<v<$ --40 \kms}
\tablenotetext{b}{measured over --40 \kms~ $<v<$ +60 \kms}
\end{tabular}
\end{center}

\tablenum{1}
\caption{Reduced $\chi^2$ over important regions of the models}

\end{table*}

\begin{figure}
\figurenum{1}
\centerline{
\psfig{figure=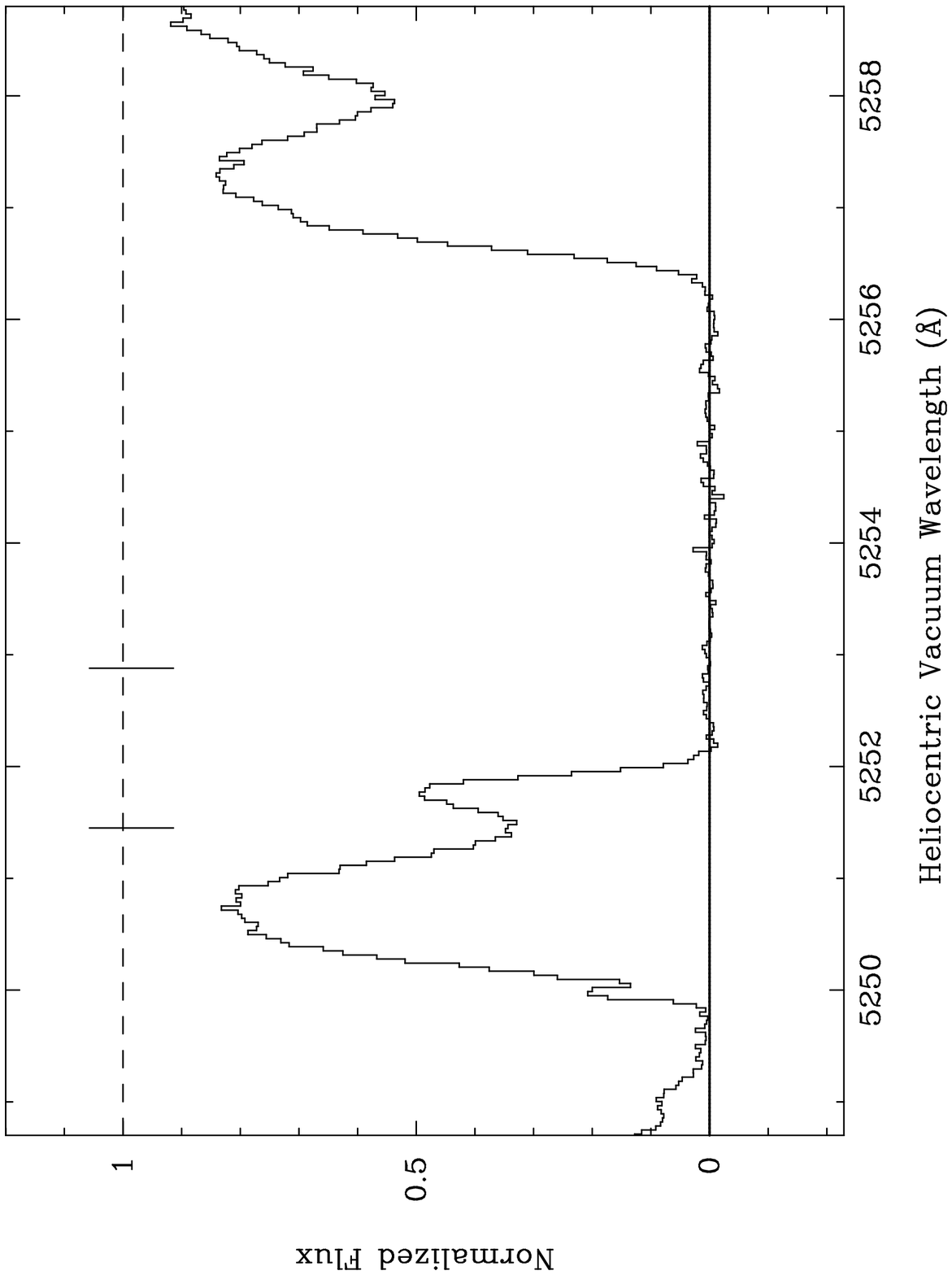,height=9.0in}}
\vspace*{-2cm}
\figcaption{}
\end{figure}
 
\begin{figure}
\figurenum{2a}
\centerline{
\psfig{figure=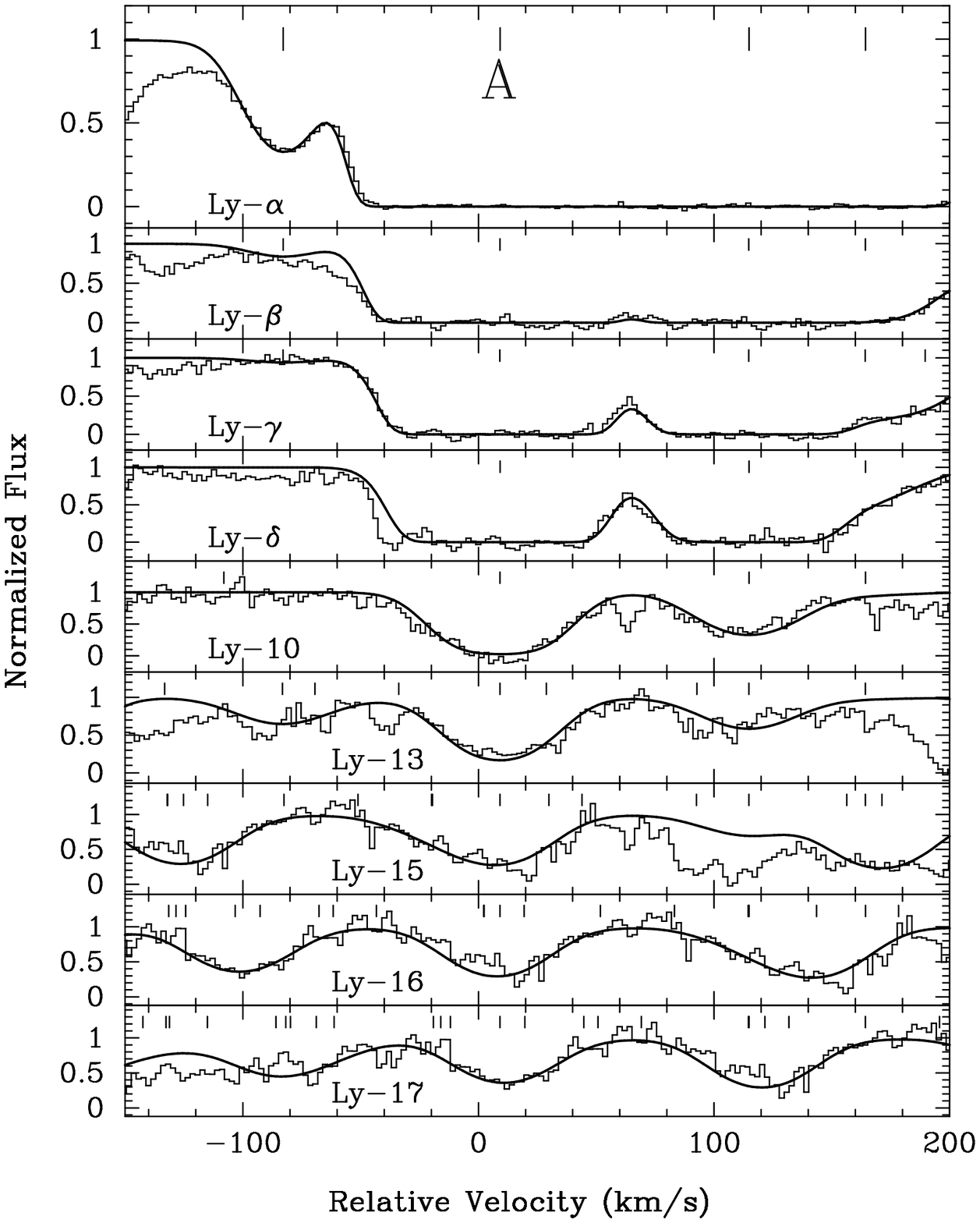,height=9.0in}}
\vspace*{-2cm}
\figcaption{}
\end{figure}
 
\begin{figure}
\figurenum{2b}
\centerline{
\psfig{figure=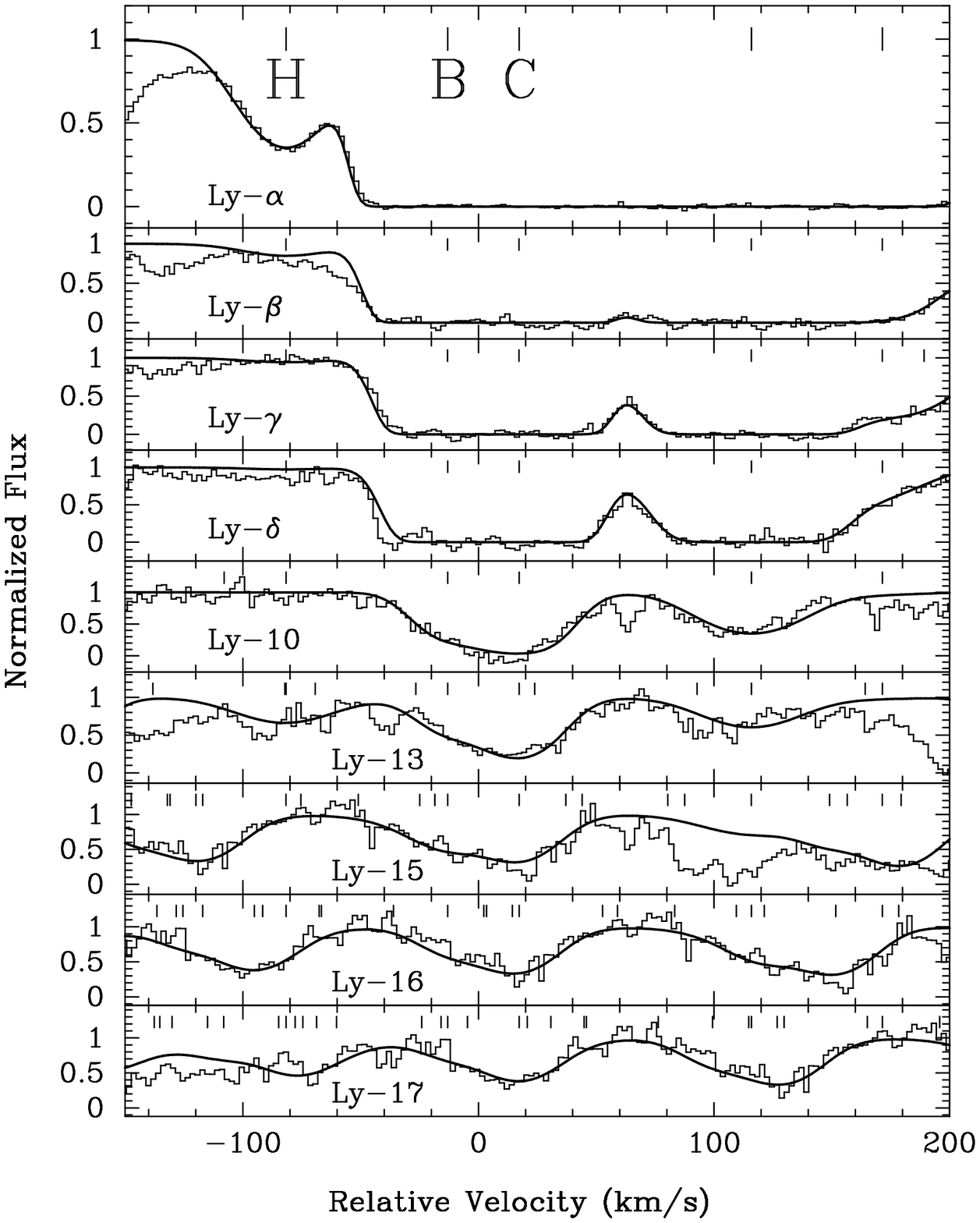,height=9.0in}}
\vspace*{-2cm}
\figcaption{}
\end{figure}
 
\begin{figure}
\figurenum{2c}
\centerline{
\psfig{figure=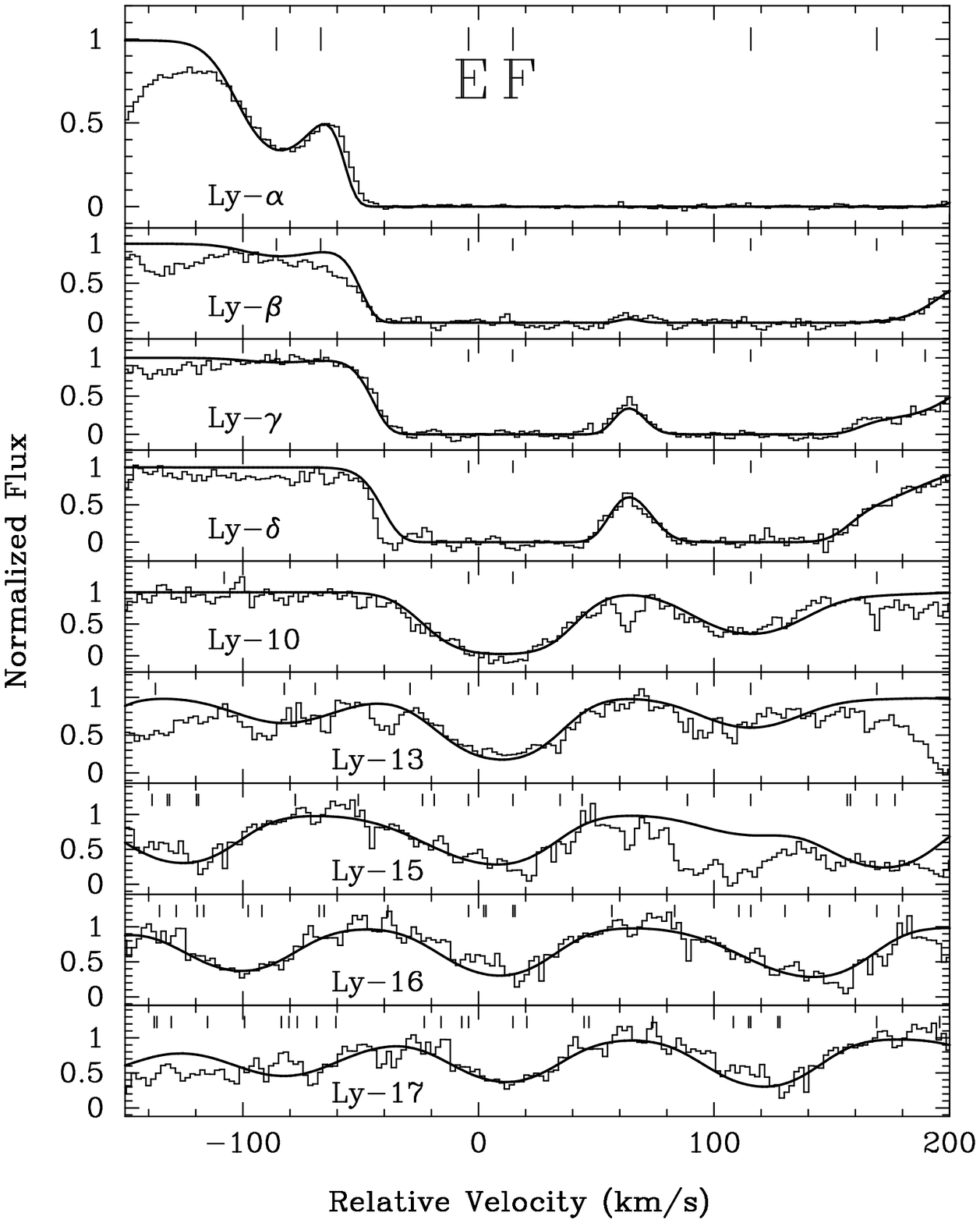,height=9.0in}}
\vspace*{-2cm}
\figcaption{}
\end{figure}
 
\begin{figure}
\figurenum{2d}
\centerline{
\psfig{figure=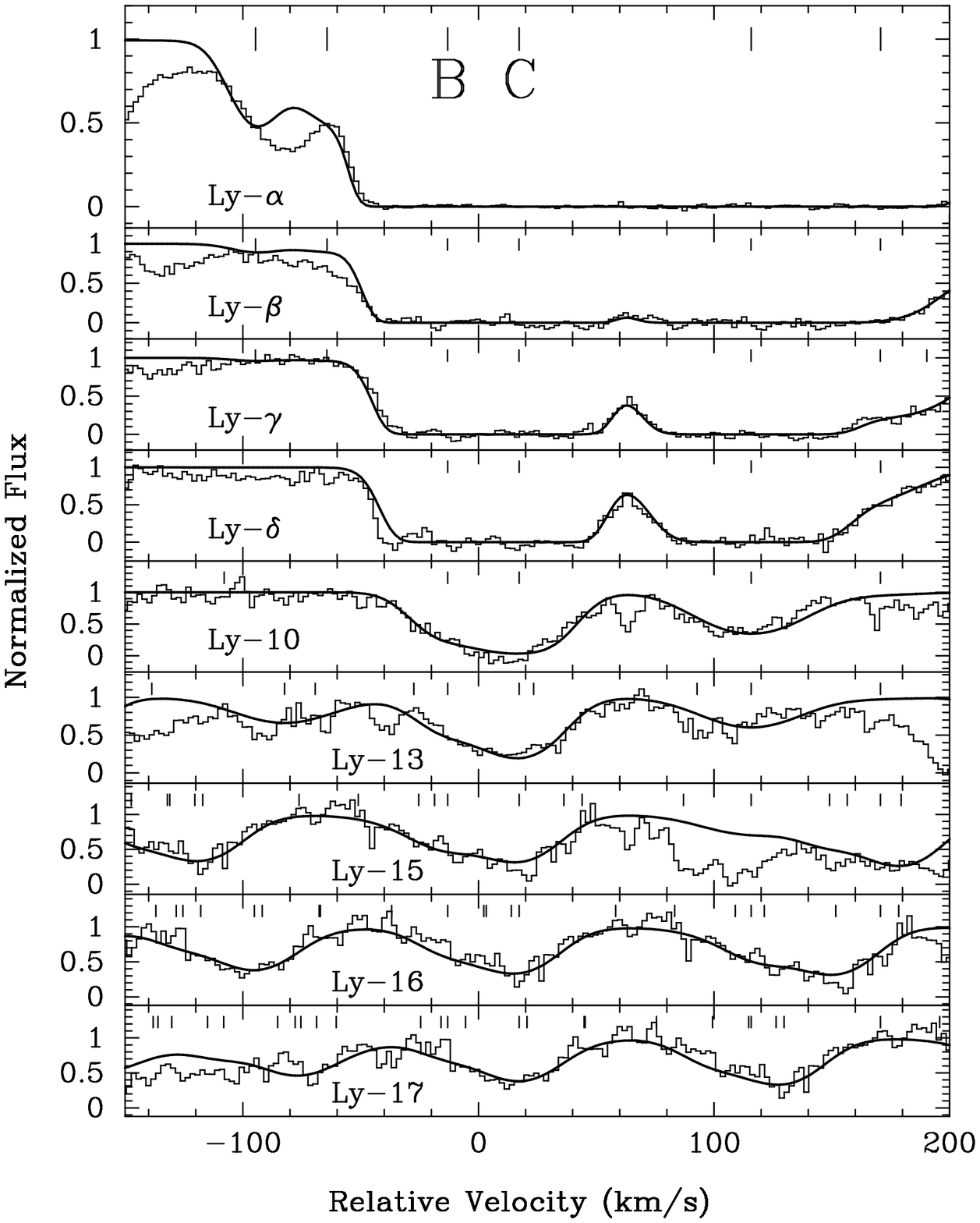,height=9.0in}}
\vspace*{-2cm}
\figcaption{}
\end{figure}

\begin{figure}
\figurenum{3a}
\centerline{
\psfig{figure=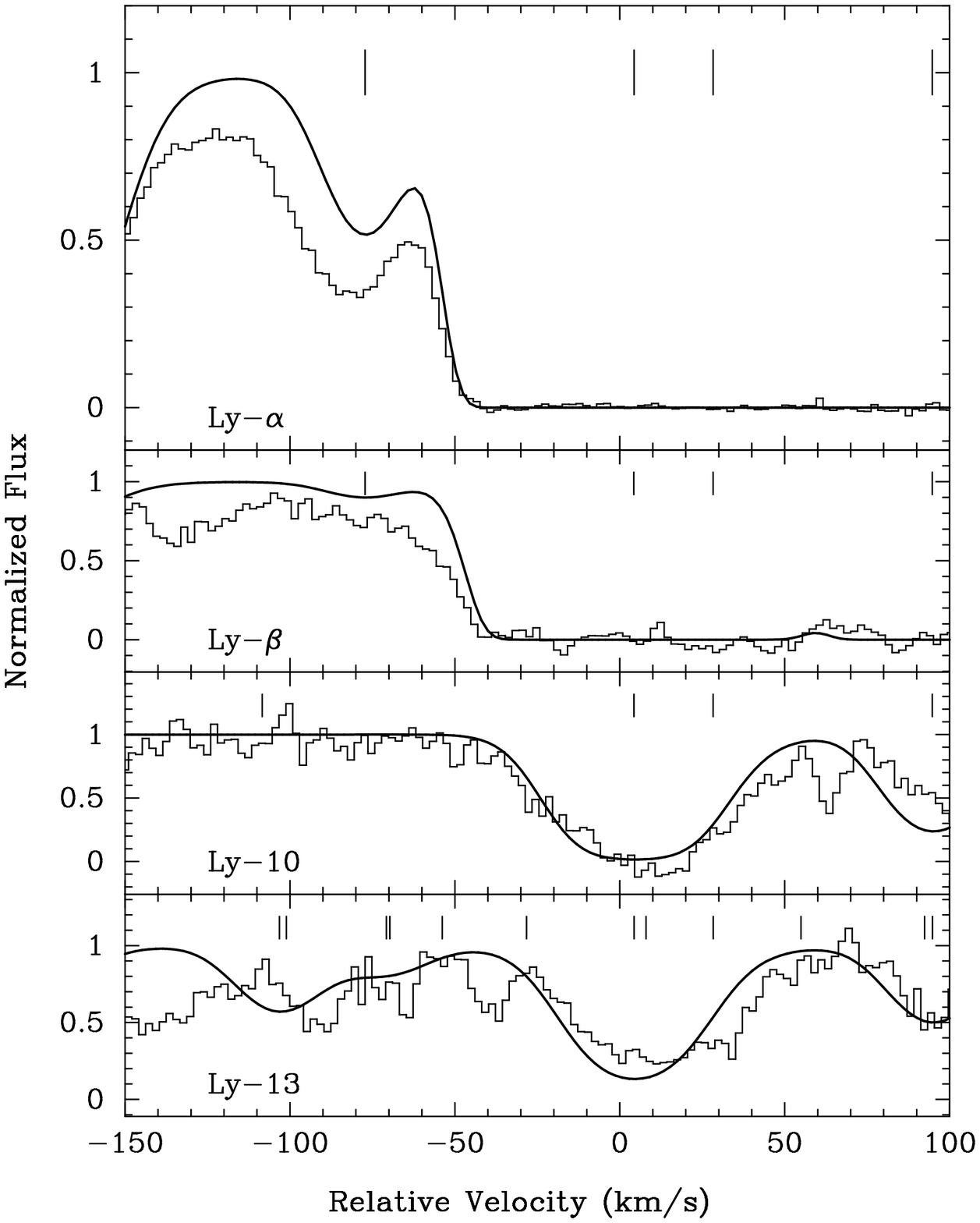,height=9.0in}}
\vspace*{-2cm}
\figcaption{}
\end{figure}

\begin{figure}
\figurenum{3b}
\centerline{
\psfig{figure=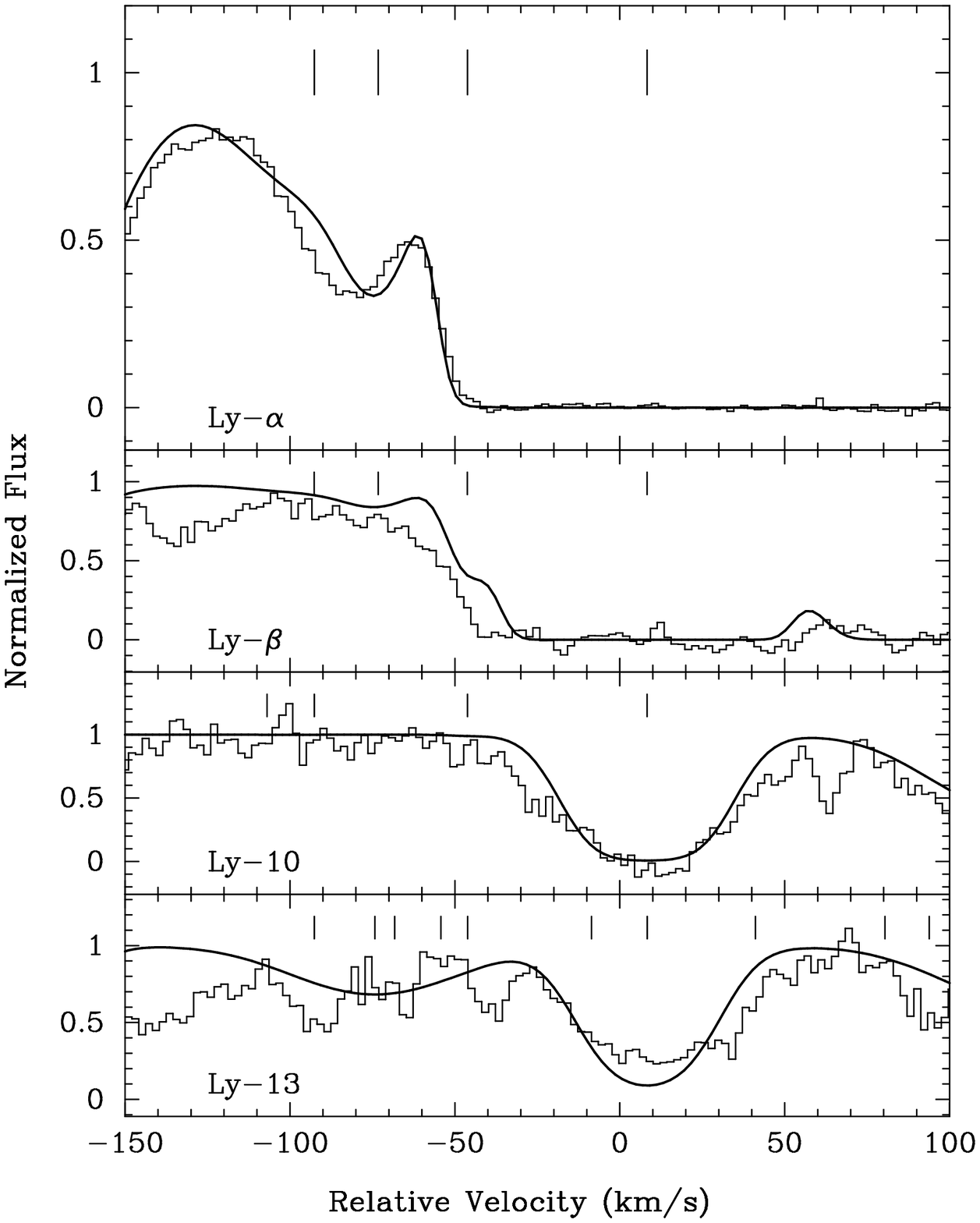,height=9.0in}}
\vspace*{-2cm}
\figcaption{}
\end{figure}

\begin{figure}
\figurenum{3c}
\centerline{
\psfig{figure=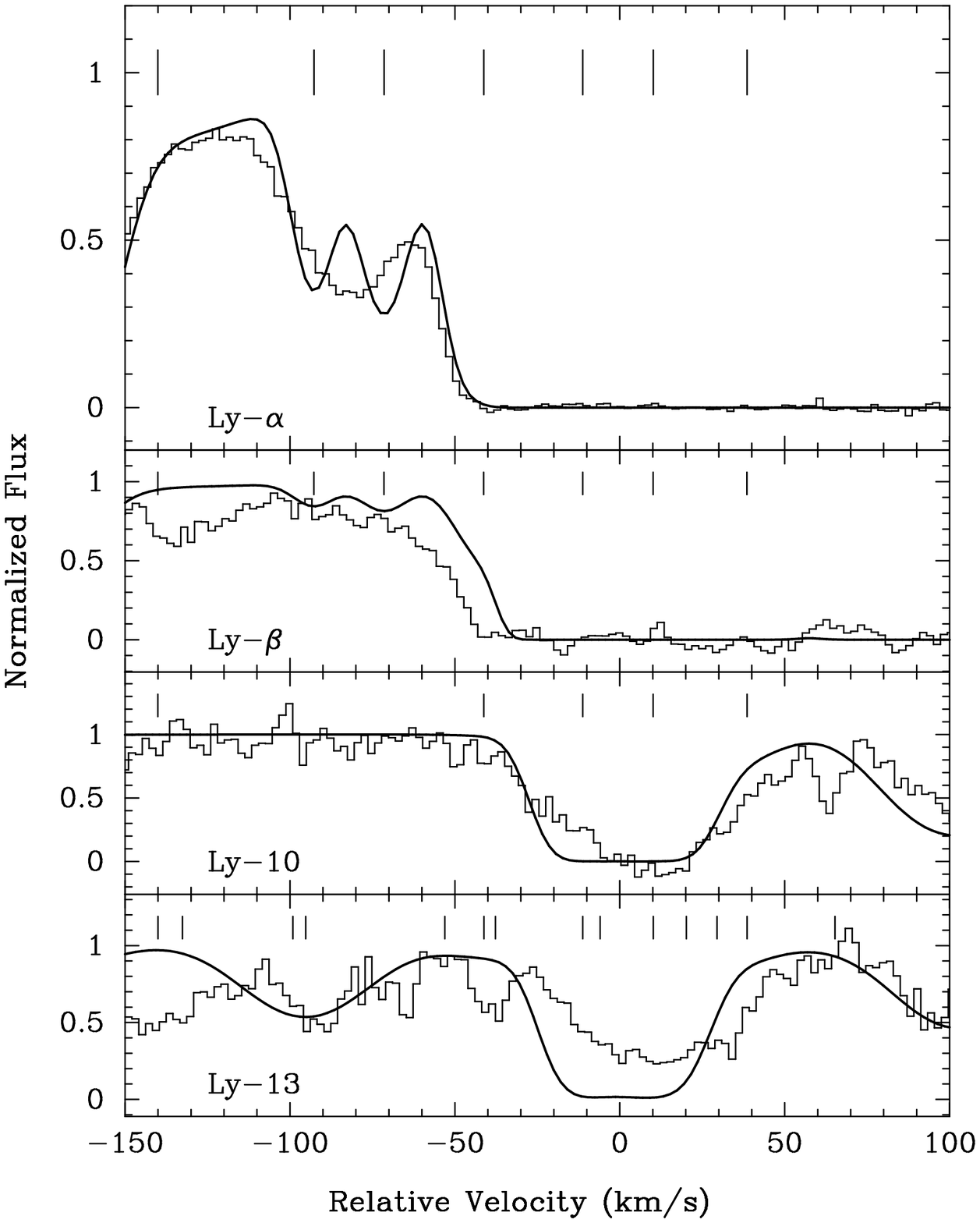,height=9.0in}}
\vspace*{-2cm}
\figcaption{}
\end{figure}

\begin{figure}
\figurenum{4}
\centerline{
\psfig{figure=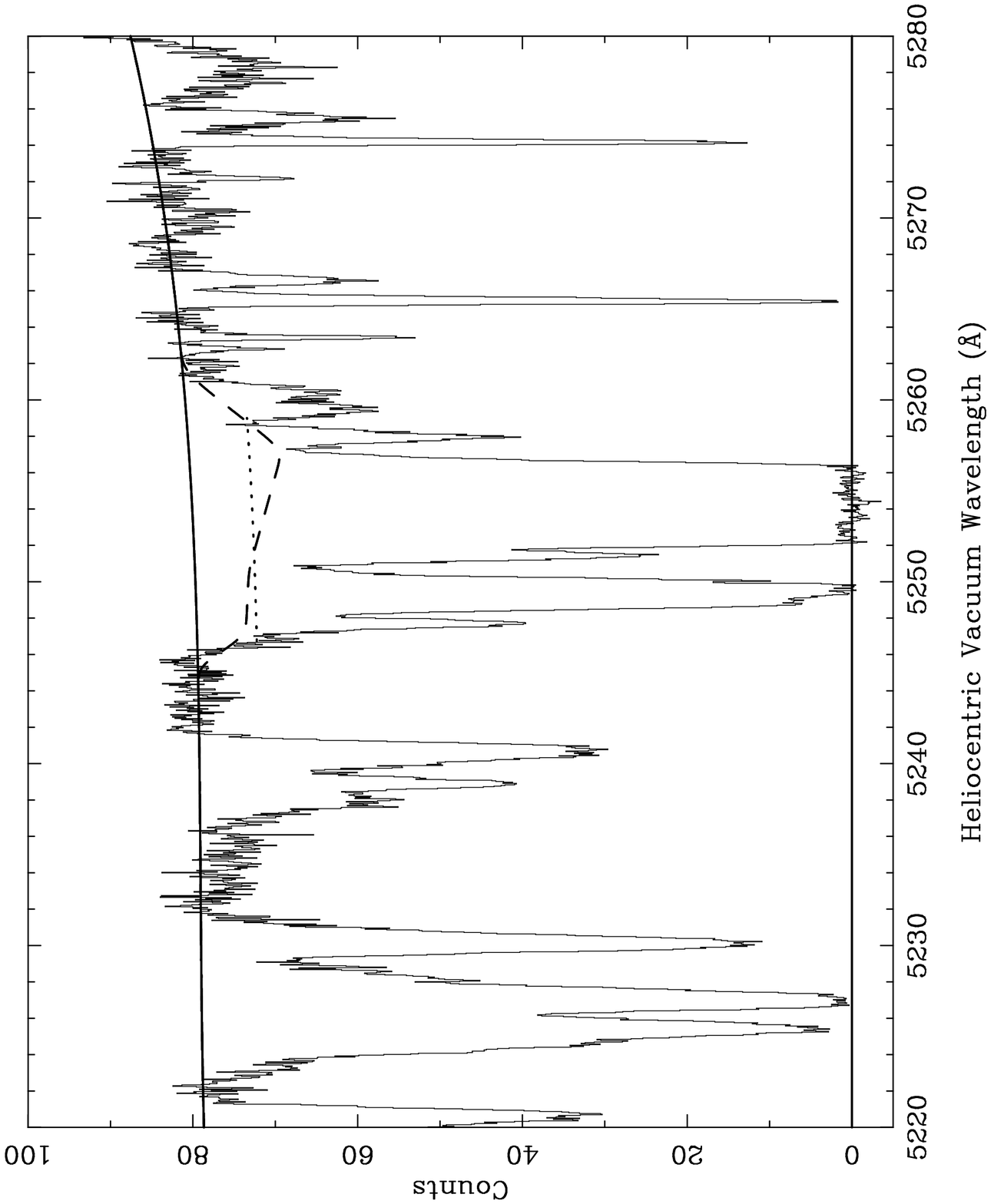,height=9.0in}}
\vspace*{-2cm}
\figcaption{}
\end{figure}

\begin{figure}
\figurenum{5}
\centerline{
\psfig{figure=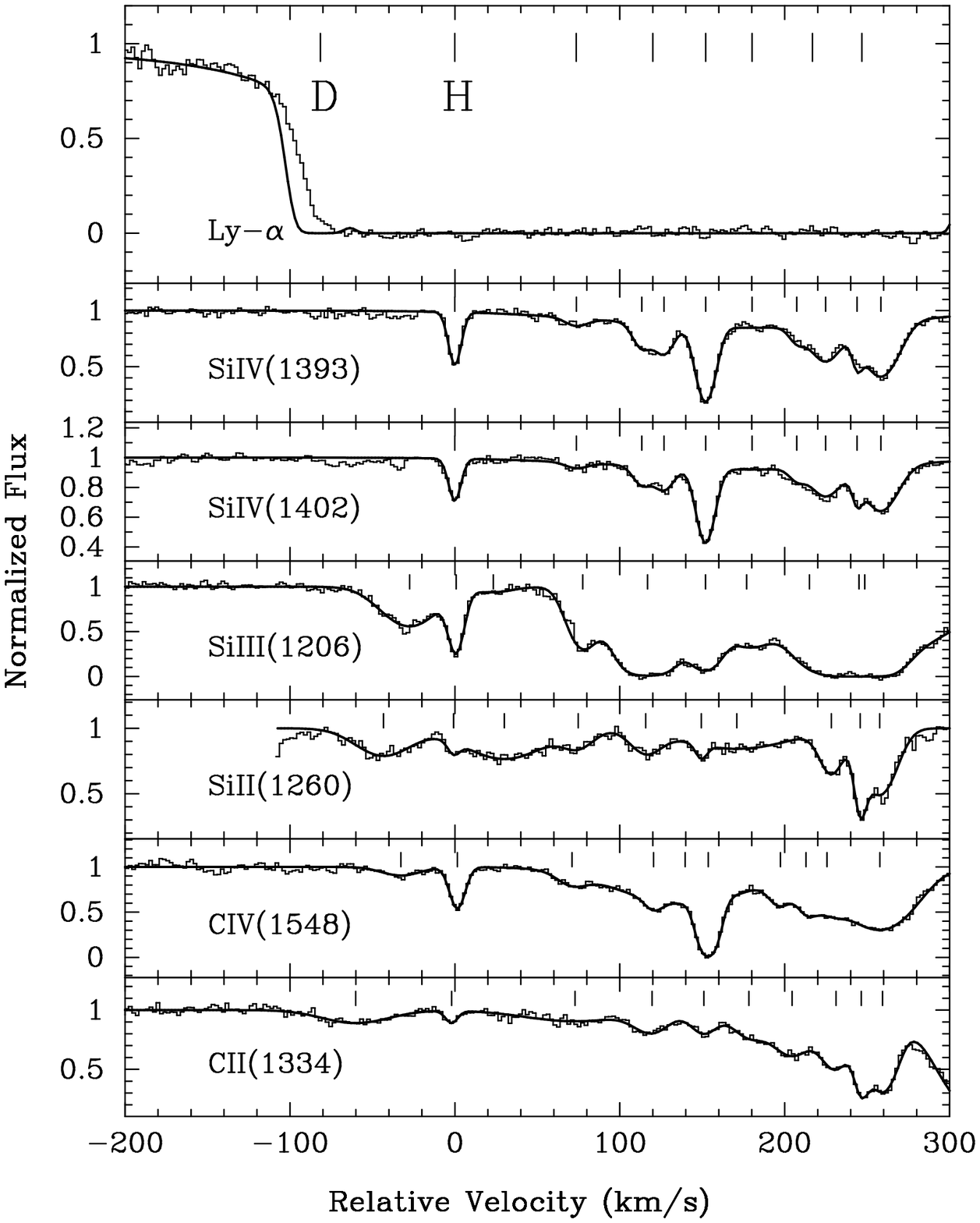,height=9.0in}}
\vspace*{-2cm}
\figcaption{}
\end{figure}

\begin{figure}
\figurenum{6a}
\centerline{
\psfig{figure=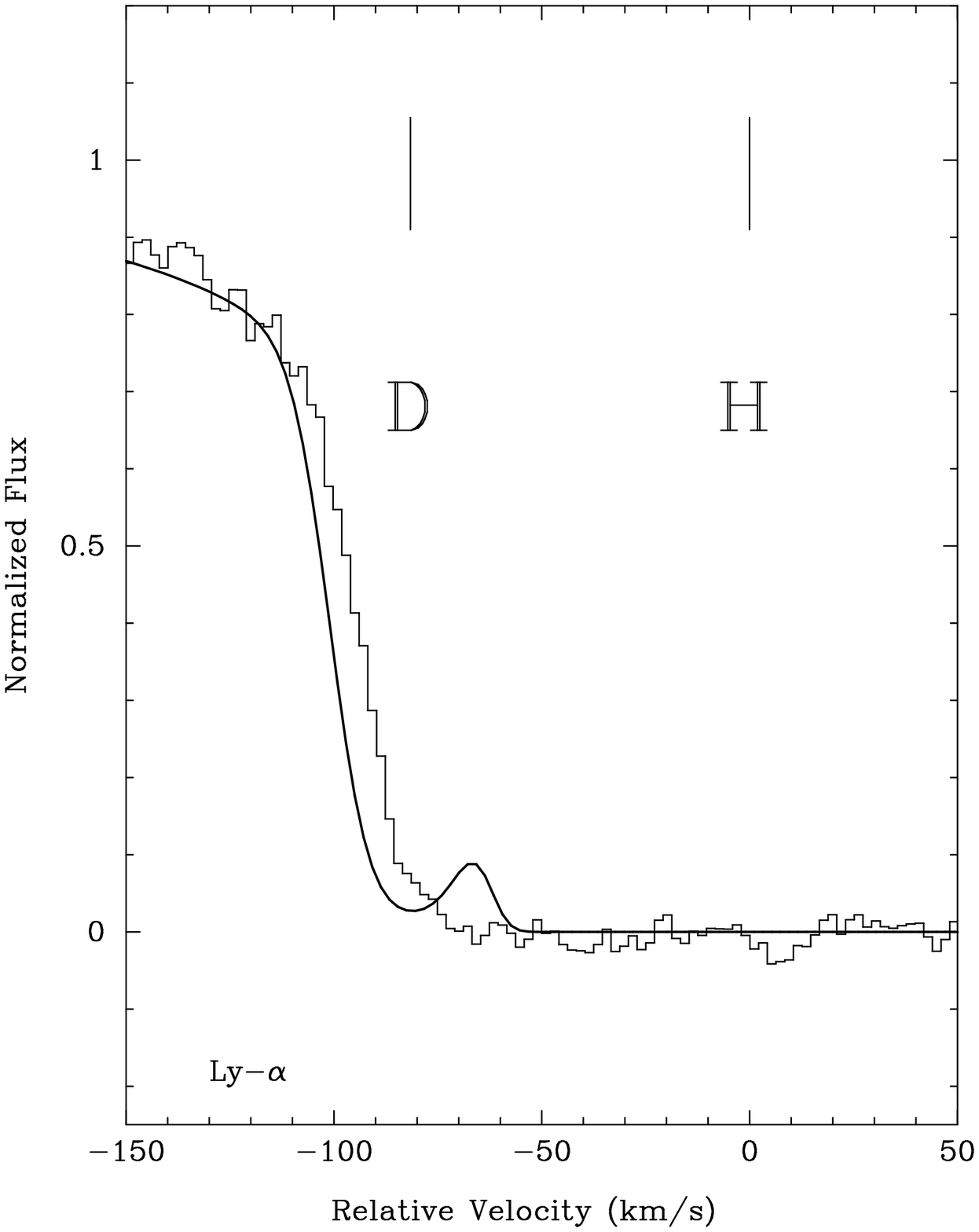,height=9.0in}}
\vspace*{-2cm}
\figcaption{}
\end{figure}

\begin{figure}
\figurenum{6b}
\centerline{
\psfig{figure=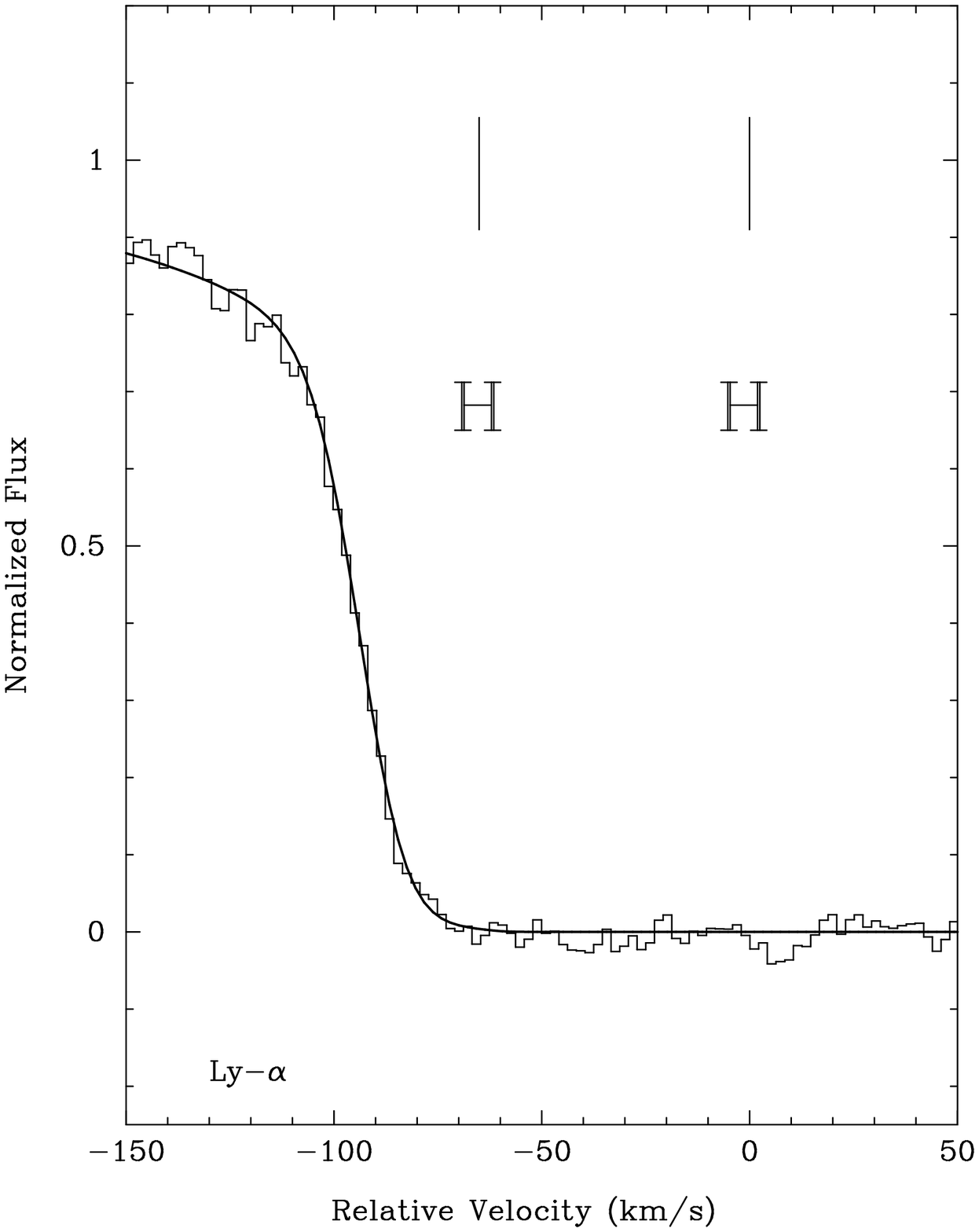,height=9.0in}}
\vspace*{-2cm}
\figcaption{}
\end{figure}

\end{document}